\documentclass[a4paper,12pt]{article}
\usepackage[mathletters]{ucs}
\usepackage[utf8x]{inputenc}

\usepackage{makeidx}
\usepackage{amsmath}
\usepackage{amssymb,latexsym}
\usepackage[normalem]{ulem}
\usepackage{cancel} 
\newtheorem{theorem}{Theorem}
\newtheorem{lemma}{Lemma}

\usepackage{url}

\title{On the Comparison of context free Grammars}
 \author{José João Almeida \and Eliana Grande \and Georgi Smirnov}
\date{~}
\begin{document}




\maketitle

\begin{abstract}
In this paper we consider the problem of context free grammars comparison
from the mathematical analysis point of view. We show that the problem can be reduced
to numerical solution of systems of nonlinear matrix equations. The
approach presented here forms a basis for comparison
algorithms oriented to automatic assessment of student’s answers
in computer science.
\end{abstract}



\section{Introduction}

In this paper we consider a language on an alphabet $V_T$, a subset of
valid words $W(V_T) = V_T^*$.
This set being usually infinite, it is necessary to use, for example, 
grammars as a mechanism for definition of the languages.
We are going to consider only
 context free grammars, covering context free languages.
The capability of writing correct grammars is an 
essential task  in computer science (used, for example, in the creation of 
programming languages, compilers, etc.).
Assessment of student's answers in computer science is a very hard and time-consuming activity. 
Computer-assisted assessment is a natural way to reduce the time spent by
the teachers in their assessment task (see, e.g.,  \cite{nelma1,nelma2}).
This paper deals with  assessment in the theory 
of context free grammars. Its main objective is to create a theoretical basis for
 algorithms allowing one to decide  if two context free
grammars are equivalent or not.
It is well-known
that the equivalence of two context free grammars is an undecidable
problem \cite{Salomaa1}. 
The problem of context free grammars equivalence was an object of
intensive studies \cite{cousot,cousot-TCS-2011}. For example, it
was solved when the equivalence is understood in structural sense
\cite{paull}, and some practical algorithms for grammars equivalence
checking were developed (see \cite{madhavan,madhavan2}, and the references
therein).

In order to present  the methodology adapt in this paper, let us consider
the following simple example.
Let the language $L$  be
$
 \{ c, ab, acb, accb, acccb, ...\}.
$
We shall denote the axiom of a grammar by $S$.
According to~\cite{Salomaa2} we can write a formal power series 
\begin{equation}
\label{s1}
 S = c + ab + acb + accb + acccb + \ldots
\end{equation}
corresponding to this language. 
 The language $L$ can be generated by the grammar
 $S \rightarrow aAb\; \mid\; c$  ;  $A \rightarrow cA\; \mid\; \epsilon $.

The following system of formal equations corresponds to this grammar
\begin{eqnarray}
&& S = aAb + c \label{s2}\\
&& A = cA  + \epsilon \label{s3}
\end{eqnarray}
Formally applying the  iteration method to this system we obtain series (\ref{s1}).
Below we define a transform that attributes a matrix meaning to formal
power series (\ref{s1}).
Namely,
\begin{itemize}
\item
any terminal letter $a,b,c$, is substituted by an 
 $({\cal N}\times {\cal N})$-matrix $\mu_a, \mu_b, \mu_c$;
\item
the nonterminal symbols $S$ and $A$ are substituted by 
 $({\cal N}\times {\cal N})$-matrix variables $S(\mu)$ and $A(\mu)$;
\item
the formal sum and product are substituted by the matrix ones;
\item
the empty word $\epsilon$  is substituted by the  $({\cal N}\times {\cal N})$ 
identity matrix $I$.
\end{itemize}
Then a matrix  $S(\mu)=S(\mu_a,\mu_b,\mu_c)$ calculated as the sum of the matrix series
$$
  S(\mu) = \mu_c 
          + \mu_a \mu_b
          + \mu_a \mu_c \mu_b
          + \mu_a \mu_c \mu_c \mu_b
          + \mu_a \mu_c \mu_c \mu_c \mu_b + ...,
$$
corresponds to $S$.
In order to effectively compute this sum we numerically solve the system
of matrix equations
 \begin{eqnarray*}
   && S(\mu) = \mu_a A(\mu) \mu_b + \mu_c \\
   && A(\mu) = \mu_c A(\mu) + I
 \end{eqnarray*}
obtained applying the transform to formal system (\ref{s2}) and (\ref{s3}).
In the same way, in general case of a grammar with the terminal alphabet
$V_T = \{a_1 , ..., a_n\}$, one can calculate the matrix
$S(\mu)=S( \mu_{a_1}, \ldots , \mu_{a_n})$.

The main result proved in this paper (Distinguishability Theorem I)
shows that if two languages  $L_1$ and $L_2$
generated by context free grammars are different and their ambiguities are bounded in some sense, then 
there exist a positive integer $\cal N$ and an $({\cal N}\times {\cal N})$-matrix substitution
$\mu_{a_1}, \ldots , \mu_{a_n}$ such that
$$ 
  {S_1}( \mu_{a_1}, \ldots , \mu_{a_n}) \neq {S_2}( \mu_{a_1}, \ldots ,\mu_{a_n}) 
$$
The condition of ambiguity boundedness excludes, for example, infinite ambiguity. 

The solution of an exercise submitted by a computer science student is supposed to be an unambiguous grammar. 
So it is very important from the point of view of automatic assessment to distinguish between
an ambiguous and an unambiguous grammars.
Note that the grammars with different ambiguities we consider as different. 


This theorem  allows us to  construct tools for  comparison
of context free grammars.
Namely, we calculate $ {S_1}( \mu_{a_1}, \ldots , \mu_{a_n})$ 
and ${S_2}( \mu_{a_1}, \ldots ,\mu_{a_n})$ for a sufficiently large
number of matrix substitutions and if   for all substitutions the 
equality ${S_1}( \mu_{a_1}, \ldots , \mu_{a_n}) = {S_2}( \mu_{a_1}, \ldots ,\mu_{a_n})$
is satisfied, then we conclude that the grammars are equivalent. 
In this paper we do not discuss the details of such algorithms, the
number of substitutions
needed to conclude that the grammars are equivalent with some probability,
etc..  This will be the subject of further research. We would like to
note that according to our experience one $(2\times 2)$ or $(3\times 3)$
random matrix substitution is enough to distinguish between two different
context free grammars.

This paper continues the research started in 
\cite{almeida_et_al2016}
where we considered  distinguishability based on $2\times 2$-matrices. Due
to negligence,  the distinguishability theorem was
not clearly formulated. We remedy this situation below (Theorem \ref{th2}).

Note also that the idea to use matrices to study formal power series is not new (cf. \cite{jungen}), 
but the approach presented in \cite{jungen} is completely  different from the one discussed here.

The paper is organized as follows. In Section 2 we prove the
distinguishability theorems. Section 3 contains convergence analysis of
the iteration method for nonlinear matrix equations. The limitations
of the method are discussed in Section 4. Section 5 contains a brief
conclusion.

\section{Distinguishability theorems}

Any context free language generated by a grammar with finite ambiguity can be defined in terms of a formal power series with
associative but not commutative variables \cite{Salomaa1,Salomaa2}.
Let  $V_T$ be the terminal alphabet, $W(V_T)$ the set of words over $V_T$, 
and $Z_+$ be the set of nonnegative integers. A map $\phi :
W(V_T)\rightarrow Z_+$
defines a formal power series
\begin{equation}
\label{eq1}
S=\sum_{P\in W(V_T)}\phi(P)P.
\end{equation}
 In the sequel, we denote the length of a word
$P\subset W(V_N\cup V_T)$ by $|P|$. The  cardinality of an alphabet $V$ is denoted by $|V|$. 
If a context free grammar is unambiguous, then the number of words of length $N$ over an alphabet
$V_T$ does not exceed $|V_T|^N$. 
We shall consider grammars generating formal power series with coefficients satisfying the following growth condition
$$
0\leq \phi (P)\leq Cq^{|P|},
$$
where $C$ and $q$ are positive constants. We shall refer to such grammars as grammars of the first class.  
All other grammars we call grammars of the second class. 
Notice that in the case of grammars of the first class, we have $\phi (\varepsilon)\leq C$.

Let $\mu$ be a map from $V_T$ to the set $R^{{\cal N}\times {\cal N}}$
of ${\cal N}\times {\cal N}$-matrices with real components. By $P(\mu)$ we will denote the
matrix obtained substituting the letters $a_i$ in $P$ by
the matrices $\mu_i$ and calculating the respective matrix product, i.e.,
if $P=a_{i_1},\ldots, a_{i_n}$, then $P(\mu)=\mu_{i_1}\ldots\mu_{i_n}$.
If the series
\begin{equation}
\label{eq2}
s(\mu)=\sum_{P\in W(V_T)}\phi(P)P(\mu)
\end{equation}
converges, its sum is an ${\cal N}\times {\cal N}$-matrix.  
Observe that the matrix series (\ref{eq2}) corresponding to a grammar of the first class converges, if
$|\mu_i|\leq M<1/(q|V_T|)$. Indeed, we have
$$
S(\mu)=\sum_{P\in W(V_T)}\phi(P) P(\mu)\leq\sum_{N\geq 0} Cq^N|V_T|^NM^N<\infty,
$$
whenever $M$ is sufficiently small. 
On the other hand, the series (\ref{eq2}) corresponding to a grammar of the second class diverges for any scalar substitution
$\mu_i=\mu >0$. Indeed, if $\phi (\varepsilon)=\infty$, 
then we have $S(\mu)\geq \phi(\varepsilon )=\infty$. 
Suppose that  $\phi (\varepsilon)\neq\infty$. 
Let $C=1$, $q=k/\mu$, $k>1$, and $\mu>0$. There exists a word  $P_k\in W(V_T)$ such that 
$$
\phi (P_k)\geq 1\cdot \left(\frac{k}{\mu}\right)^{|P_k|}.
$$
Hence we get
$$
S(\mu)=\sum_{P\in W(V_T)}\phi(P) P(\mu) \geq \left(\frac{k}{\mu}\right)^{|P_k|}\mu^{|P_k|}\geq k.
$$
Since $k$ is arbitrary, we obtain $S(\mu)=\infty$.  Thus, scalar substitutions allows us to distinguish between
grammars of the first and the second classes. We shall show that matrix substitutions permit to distinguish between 
two different context free grammars of the first class. The  distinguishability via matrix substitutions of  grammars
of the second class is not considered here.

The following
{\em distinguishability} theorems form a theoretical basis for
assessment algorithms.

\subsection{General distinguishability theorems}

\begin{theorem}[distinguishability I]
\label{th1}
Let $S_1$ and $S_2$ be two different formal power series corresponding to
context free grammars of the first class. Then there exist a positive integer $\cal N$ and
a matrix substitution  $\mu:V_T\rightarrow R^{{\cal N}\times {\cal N}}$
such that $S_1(\mu)\neq S_2(\mu)$.
\end{theorem}

To prove the theorem we need an auxiliary lemma.

\begin{lemma}
\label{lem1}
Let $U$ and $V$ be two different finite formal series composed of words of length $N$. Then
there exist a positive integer $\cal N$ and a matrix substitution
$\mu:V_T\rightarrow R^{{\cal N}\times{\cal N}}$ such that $U(\mu) \neq V(\mu)$.
\end{lemma}

\noindent{\em  Proof}.
Let $a_{j_1}\ldots a_{j_k}a_{i_1}\ldots a_{i_l}\in U \cup V$, where $l \leq N$ and $k+l=N$.
We say that $a_{i_1}\ldots a_{i_l}$ is a suffix.
Denote the set of suffixes of $U\cup V$ by $\mathcal{S}$. By $\cal N$ we denote the cardinality of $\cal S$ plus one. Let us consider the set
of unit orthogonal vectors  
$\{ e_0\} \cup \{ e_{i_1\ldots i_l} \mid a_{i_1} \ldots a_{i_l}\in\mathcal{S}\}$ 
in an $\cal N$-dimensional space. We define the linear operators $\mu_i$, $i=\overline{1,I}$ by:
$$
 \mu_i e_0 = \left\{ \begin{array}{ll}
    e_i,& a_i \in \mathcal{S} \\
    0, &\textrm{otherwise}
 \end{array} \right.
$$
$$
 \mu_i e_{i_1\ldots i_l} = \left\{ \begin{array}{ll}
e_{ii_1\ldots i_l}, & a_ia_{i_1}\ldots a_{i_l} \in \mathcal{S} \\
0, &\textrm{otherwise}\\
\end{array} \right.
$$
Let $a_{i_1}\ldots a_{i_N}\in U\cup V$ be a word. The corresponding
linear operator has the form $\mu_{i_1}\ldots \mu_{i_N}\in R^{{\cal N}\times{\cal N}}$. Obviously we have
$\mu_{i_1}\ldots \mu_{i_N}e_0=e_{i_1\ldots i_N}$. Hence we get
$$
U(\mu)e_0=\sum_{(a_{i_1}\ldots a_{i_N})\in U}e_{i_1\ldots i_N}
\neq
\sum_{(a_{i_1}\ldots a_{i_N})\in V}e_{i_1\ldots i_N}=
V(\mu)e_0,
$$
Indeed, the sets of orthogonal vectors in the two sums are different. $\Box$

\vspace{5mm}

Consider a simple example that may help the reader to better understand the construction used in the proof of Lemma \ref{lem1}. 
Let $U=a_1a_2$ and $V=a_2a_1$. In this case ${\cal N}=5$. Set
\begin{eqnarray*}
&& e_0=(1,0,0,0,0),\\
&& e_1=(0,1,0,0,0),\\
&& e_2=(0,0,1,0,0),\\
&& e_{12}=(0,0,0,1,0),\\
&& e_{21}=(0,0,0,0,1),
\end{eqnarray*}
and put
\begin{eqnarray*}
&& \mu_1e_0=e_1,\;\;\mu_1e_1=0,\;\;\mu_1e_2=e_{12},\;\;\mu_1e_{12}=0,\;\;\mu_1e_{21}=0,\\
&& \mu_2e_0=e_2,\;\;\mu_1e_1=e_{21},\;\;\mu_2e_2=0,\;\;\mu_2e_{12}=0,\;\;\mu_1e_{21}=0.
\end{eqnarray*}
Then we have
$$
\mu_1\mu_2 e_0=\mu_1e_2=e_{12}\neq e_{21}=\mu_2e_1=\mu_2\mu_1e_0.
$$
Obviously, in this example it is possible to distinguish between $U$ and $V$ with the help of $2\times 2$-matrices.
Lemma \ref{lem1} establishes only the possibility of distinguishability without calculating the minimal needed dimension $\cal N$.

\vspace{5mm}


\noindent{\em Proof of Theorem \ref{th1}}.  Since the series are
different, they admit the following representation:
$ S_1 = S_0 + U + R_1 $   and $ S_2 = S_0 + V + R_2, $
where $S_0$ is the part of coinciding words of length less than or equal
to $N$, $U$ and $V$ are different parts composed of words with length
equal to $N$, and $R_1$ and $R_2$ contain terms with words of length
greater than $N$.
By Lemma \ref{lem1} there exist a positive integer $\cal N$ and a matrix
substitution  $\mu:V_T\rightarrow R^{{\cal N}\times{\cal N}}$ such that
$U(\mu) \neq V(\mu)$. Let $t>0$.
Then we have
\begin{equation}
\label{eq10}
\Delta(t)=S_1(t\mu )-S_2(t\mu)=t^N(U(\mu)-V(\mu))+(R_1(t\mu)-R_2(t\mu)).
\end{equation}
The norms of the matrices $\mu_i$, $i=\overline{1,I}$, constructed in
Lemma \ref{lem1}, do not exceed some $\sigma>0$.
Since the grammars are of the first class, we obtain
$$
\| R_1(t\mu)\|\leq C_1\left((q_1\sigma t)^{N+1}+(q_1\sigma t)^{N+2}+\ldots\right)
$$
and
$$
 \| R_2(t\mu)\|\leq C_2\left((q_2\sigma t)^{N+1}+(q_2\sigma t)^{N+2}+\ldots\right)
$$
Let $C=\max\{C_1,C_2\}$, $q=\max\{q_1,q_2\}$ and $t<1/(q\sigma)$. Then we get
$$
\|R_1(t\mu)-R_2(t\mu)\|\leq 2C\left( (q\sigma t)^{N+1} +(q\sigma t)^{N+2}+\ldots\right)=2C(q\sigma)^{N+1}\frac{t^{N+1}}{1-q\sigma t}.
$$
From this and (\ref{eq10}) we see that $\Delta (t)\neq 0$ whenever $t>0$
is sufficiently small.  $\Box$


In many situations it suffices to consider matrix substitutions
$\mu:V_T\rightarrow R^{2\times 2}$. We associate with the symbols
$a_i\in V_T$, $i=\overline{1,I}$, pairs of variables $u_i$ and $v_i$,
$i=\overline{1,I}$.  Let
$$
U=\sum_{\{(k'^m_N,\ldots,k'^m_1)\mid m=\overline{1,M}\}}a_{k'^m_N}\ldots a_{k'^m_1}\;\;\;{\rm and}\;\;\;
V=\sum_{\{(k''^m_N,\ldots,k''^m_1)\mid m=\overline{1,M}\}}a_{k''^m_N}\ldots a_{k''^m_1}
$$
be two sets of words. Consider two sets of associated polynomials 
$$
{\cal P}_U=\left\{\prod_{j=l+1}^Nu_{k'^m_j}v_{k'^m_l} \mid l=\overline{1,N}\right\},\;\;\;{\rm and}\;\;\;
{\cal P}_V=\left\{\prod_{j=l+1}^Nu_{k''^m_j}v_{k''^m_l} \mid l=\overline{1,N}\right\}.
$$
(Here $u_{k'^m_{N+1}}=u_{k''^m_{N+1}}=1$.)
We say that $U$ and $V$ satisfy condition  $({\cal P})$ if ${\cal P}_U\neq{\cal P}_V$.

\begin{lemma}
\label{lem3}
Assume that  $U$ and $V$ satisfy condition $({\cal P})$, then there 
exists a
matrix substitution $\mu:V_T\rightarrow R^{2\times 2}$ such that 
$U(\mu) \neq V(\mu)$.
\end{lemma}

\noindent {\em Proof}. Consider the matrices 
$$
\mu_i=
\left(
\begin{array}{cc}
u_i & v_i\\
0 & 1
\end{array}
\right).
$$
By induction we easily obtain
$$
\prod_{i=1}^N\mu_{k_i}
  =\left(
\begin{array}{cc}
\prod_{i=1}^Nu_{k_i} & \sum_{i=1}^N\prod_{j=i+1}^Nu_{k_j}v_{k_i}\\
0 &1
\end{array}
\right).
$$
From this representation we see that ${\cal P}_U\neq{\cal P}_V$ implies
$U(\mu)\neq V(\mu)$. $\Box$

\begin{theorem}[distinguishability II]
\label{th2}
Let $S_1$ and $S_2$ be formal power series corresponding to context free
grammars  of the first class. Assume that the series admit the following representation:
$ S_1 = S_0 + U + R_1 $ 
and 
$ S_2 = S_0 + V + R_2,$
where $S_0$ is the part of coinciding words of length less than or
equal to $N$, $U$ and $V$ are different parts composed of words with
length equal to $N$, and $R_1$ and $R_2$ contain terms with words of
length greater than $N$. If $U$ and $V$ satisfy condition $\cal P$, then
there exists a matrix substitution  $\mu:V_T\rightarrow R^{2\times 2}$
such that  $S_1(\mu) \neq S_2(\mu)$.
\end{theorem}

\noindent \emph{Proof}. Using Lemma \ref{lem3} and following the proof of
Theorem \ref{th1} we obtain the result. $\Box$


\paragraph{Examples:}

Let $U=\{aab,bab\}$ and $V=\{aba,bba\}$. In this case condition $\cal P$
is satisfied since $u_1u_1v_2\in{\cal P}_U$ and $u_1u_1v_2\not\in{\cal P}_V$. 
On the other hand, there exist languages/grammars that cannot be
distinguished with the help of $2\times 2$-matrices. For example, using
Maxima computer algebra system it is easy to show that for any choice of
$2\times 2$-matrices $\mu_1$ and $\mu_2$ we have
$$ 
\mu_1 \mu_1 \mu_2 \mu_2 \mu_1+\mu_1 \mu_2 \mu_1 \mu_1 \mu_2+\mu_2 \mu_1 \mu_2 \mu_1 \mu_1= \mu_1 \mu_1 \mu_2 \mu_1 \mu_2+\mu_1 \mu_2 \mu_2 \mu_1 \mu_1+\mu_2 \mu_1 \mu_1 \mu_2 \mu_1.
$$
Therefore the languages
\begin{equation}
\label{ll}
\{ aabba, abaab, babaa\}\;\;{\rm and}\;\; \{aabab, abbaa, baaba\}
\end{equation}
cannot be distinguished using $2\times 2$-matrices. However substituting
$3\times 3$-matrices it is easy show  that the languages (\ref{ll})
are different.

\subsection{Comparison of short words}

In many situations the difference between two grammars can be detected
comparing short words trough substitution of $2\times 2$-matrices.
We fulfilled about $47\cdot 10^6$ tests to analyze all finite languages 
over terminal alphabet $\{ a,b,c\}$  containing no more than three 
words of length less than or equal to five. We found that only  the 
following languages cannot be distinguished using $2\times 2$-matrix 
substitutions. Namely, the pair of the  languages 
\begin{equation}
\label{L}
S_{1} : a a b c a | a b a a c | b a c a a \;\;\; {\rm and}\;\;\;
S_{2} : a a b a c | a b c a a | b a a c a  
\end{equation}
and other  pairs obtained as the result of permutation of the letters 
$\{ a,b,c\}$ and/or substitution of $c$ by $a$ or $b$. We shall denote
this set of pairs of languages by $L$

This proves the following theorem.
\begin{theorem}[distinguishability III]
\label{th3}
Let $S_1$ and $S_2$ be formal power series corresponding to context free
grammars  of the first class over the terminal alphabet $\{ a,b,c\}$ and let $N\leq 5$. Assume
that the series admit the following representation:
$ S_1 = S_0 + U + R_1 $
and 
$ S_2 = S_0 + V + R_2, $
where $S_0$ is the part of coinciding words of length less than or equal
to $N$, $U$ and $V$ are different parts composed of no more than three
words with length equal to $N$, and $R_1$ and $R_2$ contain terms with
words of length greater than $N$. If the pair $U$ and $V$ does not coincide
with one of the pairs from $L$, then there exists a matrix substitution
$\mu:V_T\rightarrow R^{2\times 2}$ such that  $S_1(\mu) \neq S_2(\mu)$.
\end{theorem}

\noindent \emph{Proof}. Following the proof of Theorem \ref{th1} we obtain the result. $\Box$


Note that the pair of languages (\ref{L}) can be used to construct
examples of infinite languages that cannot be distinguished using 
$2\times 2$-matrices. One of such examples is the pair
$$ S \rightarrow  a a b A a  \mid  a b a a A  \mid  b a A a a 
\;\; ; \;\; 
   A \rightarrow a A  \mid   b
$$
and
$$ S  \rightarrow   a a b a A  \mid  a b A a a  \mid  b a a A a
\;\; ; \;\;
   A  \rightarrow   a A  \mid   b
$$ 
Note also that in all the grammars of several programming languages and
of educational exercises that we analyzed, the distinguishability was
always possible using $(2\times 2)$-matrices.

\section{Systems of nonlinear matrix equations}

It is well-known \cite {Salomaa1,Salomaa2} that to any context free
grammar there corresponds a system of nonlinear equations that allows
one to obtain the respective formal power series via successive
iterations. The terms of the series are the words of the respective
language. This correspondence between series and systems of nonlinear
equations makes it possible to effectively compute the sums of the series
for any ${\cal N}\times{\cal N}$-matrix substitution.
Let $X_i$, $i=\overline{0,m}$, be the nonterminals of a context free
grammar and let  $P^i_{j}$, $i=\overline{0,m}$, $j=\overline{1,l_i}$,
be the words that appear in the right-hand sides of productions with
the left-hand sides $X_i$. The system of equations corresponding to the
grammar has the form
\begin{eqnarray}
&& X_1=P^1_1+\ldots+P^1_{l_1},\nonumber\\
&& \hspace{20mm}\vdots \label{7}\\
&& X_m=P^m_1+\ldots+P^m_{l_m}.\nonumber
\end{eqnarray}
Substituting the symbols of the terminal alphabet $a_k\in P^i_j$ by
matrices $\mu_k$, we obtain a system of nonlinear matrix equations with
unknowns $X_i$. This system, $X=F(X)$, can be solved using the iterative
process $X^{k+1}=F(X^k)$, $X^0=0$, or using the Newton method. As we
shall see in the sequel the convergence of the method of successive
iterations can be guaranteed for a large class of grammars, for example
for the grammars in Chomsky and Greibach normal forms. Note that for
regular grammars system (\ref{7}) is linear.

\subsection{Convergence of successive iterations}

Let us consider a context free grammar with productions
$$ X_i\rightarrow \mathcal{P}_j^{0,i}, \;\; i=\overline{1,n},\; j=\overline{1,J_i^{\cal P}}
\;\; ; \;\;
X_i\rightarrow p_j^i, \;\; i=\overline{1,n},\; j=\overline{1,J_i^{p}},
$$
where $\mathcal{P}_j^{0,i}\in W(V_N \cup V_T) \setminus  W(V_T)$  and
$p_j^i\in W(V_T)$, $p_j^i\neq\varepsilon$.
We assume that the words ${\cal P}_j^i$ contain more than one
symbol. (For example, any grammar without renaming and $\varepsilon$-rules is of this type.) The structure of these words can be described in the
following manner:
$
{\cal P}_j^{l,i}=q^{l+1,i}_jX_{k^{l+1,i}_j} {\cal P}_j^{l+1,i},\;\;\; l=\overline{1,L_j^i},
$
where ${\cal P}_j^{l,i}\in W(V_N\cup V_T)\cup\{\varepsilon\}$ and
$q_j^{l,i}\in W(V_T)\cup\{\varepsilon\}$.
The corresponding system of equations has the following structure:
\begin{eqnarray}
\label{c1}
&&  X_i = F_i(X_1,X_2,\ldots,X_n) = \sum_j\mathcal{P}_j^{0,i} + \sum_j p_j^i 
\end{eqnarray}
To simplify the notations we denote by $P$ the matrix $P(\mu)$ obtained
substituting the symbols $a_i$ and $X_i\in P$ by the matrices $\mu_i$
and $\xi_i$, respectively. We use the notation $\tilde{P}$ when the
symbols $X_i\in P$ are substituted by the matrices $\tilde{\xi}_i$. Let
$X=(X_1,\ldots,X_n)$  and
$\tilde{X}=(\tilde{X}_1,\ldots,\tilde{X}_n)$ be two collections of $n$
matrices ${\cal N}\times{\cal N}$. Then, using our notations, we have
$$
{\cal P}_j^{0,i}-\tilde{\cal P}_j^{0,i}=q_j^{1,i}(X_{k^{1,i}_j}{\cal P}_j^{1,i}-\tilde{X}_{k^{1,i}_j}\tilde{\cal P}_j^{1,i})
 =q_j^{1,i}(X_{k^{1,i}_j}-\tilde{X}_{k^{1,i}_j}){\cal P}_j^{1,i}+\tilde{X}_{k^{1,i}_j}({\cal P}_j^{1,i}-\tilde{\cal P}_j^{1,i})
$$
$$
 =q_j^{1,i}(X_{k^{1,i}_j}-\tilde{X}_{k^{1,i}_j}){\cal P}_j^{1,i}+\tilde{X}_{k^{1,i}_j}
(q_j^{2,i}(X_{k^{2,i}_j}-\tilde{X}_{k^{2,i}_j}){\cal P}_j^{2,i}+\tilde{X}_{k^{2,i}_j}({\cal P}_j^{2,i}-\tilde{\cal P}_j^{2,i}))
$$
\begin{equation}
\label{YZ}
 =\ldots=Y_j^{1,i}(X_{k^{1,i}_j}-\tilde{X}_{k^{1,i}_j})Z_j^{1,i}+\ldots+Y_j^{n,i}(X_{k^{n_j^i,i}_j}-\tilde{X}_{k^{n_j^i,i}_j})Z_j^{n,i},
\end{equation}
where $Y_j^{l,i},Z_j^{l,i}\in W(V_N\cup V_T)$ and $n_j^i$ is the number
of nonterminals in the word ${\cal P}_j^{0,i}$. Assume that the norms of
all matrices $a_i$, $X_i$, and $\tilde{X}_i$ do not exceed  $\delta >0$
and that $\bar{n}\delta<1$, where $\bar{n}=\sum_{i,j}n_j^i$.
Then from the representation  (\ref{YZ}) we obtain
\begin{equation}
\label{qq}
\max_{i=\overline{1,n}}\|F_i(X)-F_i(\tilde{X})\| \leq \bar{n}\delta\max_{j=\overline{1,n}}\| X_j - \tilde{X_j} \|
\end{equation}
Set $\tilde{X}=0$. Then if $\| X_i\|<\delta$, we get
$$
\max_{i=\overline{1,n}}\|F_i(X)\|\leq \bar{n}\delta \max_{j=\overline{1,n}}\|X_j\|<\bar{n}\delta^2<\delta.
$$
Let us consider a closed ball 
${\cal B}=\{X\mid \max_{j=\overline{1,n}}\|X_j\|\leq \delta\}$ in the
space of  matrices $X=(X_1,\ldots,X_n)$. We have proved that 
$F({\cal B})\subset{\cal B}$ and that $F$ is a contracting map. Hence 
there exists
a unique fixed point $\hat{X}=F(\hat{X})\in {\cal B}$.  This fixed point
is the limit of the sequence of iterations
\begin{equation}
\label{Xk}
X^{k+1}=F(X^k),\;\;\; k=0,1,\ldots,\;\; X^0=0.
\end{equation}
Thus we have the following result.

\begin{theorem}[Convergence]
\label{thconv}
Assume that the system of nonlinear equations corresponding to a
context free grammar has form (\ref{c1}) and that the words with
nonterminal symbols, ${\cal P}_j^i$, contain more than one symbol. Then
substituting the terminal symbols by matrices with a sufficiently
small norm (less than $\delta$), we can guarantee the convergence of
the sequence
 (\ref{Xk}) to a unique solution of the matrix system (\ref{c1}).  
\end{theorem}

\paragraph{Example:}

Let us consider the grammar
\begin{equation}
\label{c3}
S\rightarrow  SaA  \mid a 
\;\; ; \;\;
A\rightarrow  cSd  \mid b
\end{equation}
The respective system of equations is
\begin{equation}
\label{c4}
\begin{array}{l}
S= F_S=SaA+a,\\
A= F_A=cSd+b.
\end{array}
\end{equation}
The conditions of Theorem \ref{thconv} are satisfied (the words $SaA$
and $cSd$ contain more than one symbol). Therefore the system can be
solved using the  iteration  method whenever the symbols  $a$, $b$, $c$,
and $d$ are replaced by matrices with a sufficiently small norm.

\subsection{Other cases where the iteration  method can be used}

In many cases the grammar may have productions of the form $A\rightarrow B$.
The iteration  method can be applied also to the corresponding system
of equations after some transformation. Namely, assume that the system
has the form
\begin{equation}
\label{eq11X}
 X = F(X) + \Lambda X,
\end{equation}
where $F$ has form (\ref{c1}) considered above and 
$\Lambda:R^{({\cal N}\times{\cal N})n}\rightarrow R^{({\cal N}\times{\cal N})n}$
is a linear operator such that there exists the inverse
$(I - \Lambda)^{-1}$. Then system (\ref{eq11X}) is equivalent  with the system
$
 X = (I -\Lambda)^{-1}F(X).
$
Obviously the map $X\rightarrow (I -\Lambda)^{-1}F(X)$ is contracting
and transforms  
${\cal B}=\{X\mid \max_{j=\overline{1,n}}\|X_j\|\leq \delta\}$ into $\cal B$,
whenever the terminal symbols are replaced by
matrices with a sufficiently small norms.

\paragraph{Example:}

Let us consider the grammar
\begin{equation}
\label{8}
S \rightarrow  SaA  \mid   A
\;\; ; \;\;
A \rightarrow  cSd  \mid   b
\end{equation}
which is a correct solution of an exercise. (This example is taken from
\cite{almeida_et_al2016}.)
The corresponding system of equations reads:
\begin{equation}
\label{e1}
\begin{array}{l}
S= SaA+A,\\
A= cSd+b.
\end{array}
\end{equation}
Below we present three other possible answers.

\paragraph*{Alternative correct solution}

The following grammar is different but generates the same language:
\begin{equation}
\label{9}
S  \rightarrow  AaS  \mid   A
\;\; ; \;\;
A  \rightarrow  cSd  \mid   b
\end{equation}
The corresponding system of equations is
\begin{equation}
\label{e2}
\begin{array}{l}
S= AaS+A,\\
A= cSd+b.
\end{array}
\end{equation}

\paragraph*{Wrong answer}

The following grammar does not generate the same language (does not
generate the word $\emph{cbabd}$):
\begin{equation}
\label{10}
S  \rightarrow  SaA  \mid   A
\;\; ; \;\;
A  \rightarrow  cAd  \mid   b
\end{equation}
The corresponding system is
\begin{equation}
\label{e3}
\begin{array}{l}
S= SaA+A,\\
A= cAd+b.
\end{array}
\end{equation}

\paragraph*{Ambiguous grammar}

The following grammar generates the same language but is ambiguous  (the word $baba$ 
can be generated by different ways):
\begin{equation}
\label{11}
S  \rightarrow  SaS \mid   A
\;\; ; \;\;
A  \rightarrow  cSd \mid   b
\end{equation}
The corresponding system has the form
\begin{equation}
\label{e4}
\begin{array}{l}
S= SaS+A,\\
A= cSd+b.
\end{array}
\end{equation}


System (\ref{e1}) corresponding to grammar (\ref{8}) is equivalent to the system
$$
\left(
\begin{array}{c}
S\\
A
\end{array}
\right)
  =
\left(
\begin{array}{cc}
I & I\\
0 & I
\end{array}
\right)^{-1}
\left(
\begin{array}{c}
SaA\\
cSd+b
\end{array}
\right).
$$
Replacing the symbols  $a$, $b$,
$c$, and  $d$ by $(2 \times 2)$-matrices with sufficiently small 
norm we get a system
that can be solved using the  iteration method. In the same way we can
transform and solve other systems.
Starting the iterative process with $S=A=0$ and solving systems
(\ref{e1}), (\ref{e2}), (\ref{e3}), and (\ref{e4})  we see that the
difference between  $S$ components of solution of systems (\ref{e1})
and (\ref{e2}) is zero, while for the pair (\ref{e1}) and  (\ref{e3})
or (\ref{e1}) and (\ref{e4}) the difference is not zero.
This allows one to clearly distinguish between right and wrong answers.

Note that the interval where the components of the matrices are generated
must be (a) sufficiently small in order to guarantee the convergence
of the iterative process, (b) big enough to distinguish between two
different languages.


Another important case deals with the grammars having productions of the
form $A\rightarrow\varepsilon$. In this case some equations have the form
$$
 X_i = F_i(X_1,X_2,\ldots,X_n) + I
$$
Introducing new variables $Y_i=X_i-I$ in many situations it is possible
to transform the system to a system satisfying conditions of Theorem
\ref{thconv}.

\paragraph{Example:}

Let us consider the grammar
$$
S\rightarrow SaA  \mid b 
\;\; ; \;\;
A\rightarrow cSd  \mid \varepsilon
$$
The corresponding system of nonlinear equations is
\begin{eqnarray*}
&& S=SaA+b\\
&& A=cSd+I
\end{eqnarray*}
Introducing new variable $B=A-I$ we obtain the system
\begin{eqnarray*}
&& S=SaB +Sa +b\\
&& B=cSd
\end{eqnarray*}
which can be solved by  iteration method.

\section{Comparison with a simple heuristic method}

One natural heuristic method to compare two context free grammars
is:
\begin{itemize}

\item Explicitly check with a parsing algorithm (for example, the
classic CYK algorithm \cite{cyk} or its modification \cite{cyk-de}) 
all words up to length $n$;

\item If difference is found, output "grammars are different";

\item If no difference is found, output "grammars could be the same".
\end{itemize}

Such an approach does not allow us to distinguish between two grammars
with different ambiguities and cannot be applied to grammars with
distinguishing words of big length.

For example, let us consider two grammars:
$$
S → \varepsilon \mid I\; I\; S \mid A \;\; ;\; \; 
I → a \mid b \mid c \mid d  \;\; ;\; \; 
A → C\; C\; C\; a\; a\; a   \;\; ;\; \; 
C → a\; a\; a\; a  
$$
and
$$
S → \varepsilon \mid I\; I\; S \mid B \;\; ;\; \; 
I → a \mid b \mid c \mid d  \;\; ;\; \; 
B → C\; C\; C\; b\; b\; b   \;\; ;\; \; 
C → b\; b\; b\; b  
$$

The productions
 $S → \varepsilon \mid I\; I\; S$ and $I → a \mid b \mid c \mid d $
generate all the words of even length containing $a$, $b$, $c$ and $d$,
while $A$ generates the word $a^{15}$ and
$B$ generates the word $b^{15}$.

Then distinguishing words exist, but have length 15 or more. 
This means that any generation and testing based algorithm
must test about $4^{15} \approx 10^9$ words. This is would be a very
time-consuming procedure.
 
The algorithm presented here can easily distinguish between
these to grammars within a fraction of a second.

\section{Limitations of iteration  method}

In some cases the  iteration  method cannot distinguish between two
different grammars. This happens when we consider grammars with very
long words from $W(V_T)$. The point is that the computer precision is
not sufficient to correctly compute products of many small numbers.
Consider the following grammar
\begin{equation}
\label{r3}
S \rightarrow A S  \mid  B S  \mid  B    \;\; ; \;\;
A \rightarrow a_1 a_2\ldots a_n          \;\; ; \;\;
B \rightarrow  a_1| a_2 | \ldots  | a_n.
\end{equation}
The corresponding system of matrix equations reads
\begin{equation}
\label{r4} 
\begin{array}{l}
S=AS + BS + B,\\
A=a_1a_2 \ldots a_n, \\
B=a_1+a_2\ldots+a_n. \\
\end{array}
\end{equation}
Let the second grammar be
\begin{equation}
\label{r31}
S  \rightarrow  A S  \mid  B S  \mid B \;\; ; \;\;
A  \rightarrow  a_2 a_1\ldots a_n      \;\; ; \;\;
B  \rightarrow  a_1| a_2 | \ldots  | a_n .
\end{equation}
with the corresponding system
of equations
\begin{equation}
\label{r41} 
\begin{array}{l}
S=AS + BS + B,\\
A=a_2a_1 \ldots a_n, \\
B=a_1+a_2\ldots+a_n. \\
\end{array}
\end{equation}
Systems (\ref{r4}) and (\ref{r41}) are equivalent to the equations  
$$
S = a_1a_2 \ldots a_nS + (a_1+a_2+\ldots+a_n)S + (a_1+a_2+\ldots+a_n)
$$ 
and 
$$
S = a_2a_1 \ldots a_nS + (a_1+a_2+\ldots+a_n)S + (a_1+a_2+\ldots+a_n),
$$
respectively. To guarantee the convergence of iterations we have to
impose the restriction
$
\|a_1\|+\|a_2\|+\ldots+\|a_n\|=\alpha<1.
$
From the inequality of arithmetic and geometric means we get
$$
\sqrt[n]{\|a_1 \|\ldots \|a_n\|} \leq (\|a_1\|+\ldots+\|a_n\|)/n.
$$
Hence for sufficiently large $n$ we have
$
\|a_1 \|\ldots \|a_n\| \leq \alpha^n / n^n < \delta, 
$
where $\delta>0$ is the computer precision. 
Thus the computer interprets the iterative processes
$$
S_{k+1} = a_1a_2 \ldots a_nS_k + (a_1+a_2+\ldots +a_n)S_k + (a_1+a_2+\ldots +a_n)
$$
and
$$
S_{k+1} = a_2a_1 \ldots a_nS_k + (a_1+a_2+\ldots +a_n)S_k + (a_1+a_2+\ldots +a_n)
$$
as the same process
$$
S_{k+1} = (a_1+a_2+\ldots +a_n)S_k + (a_1+a_2+\ldots +a_n)
$$
and, therefore, the method does not allow to distinguish between two
grammars. 

However, the method can be applied to \emph{real size 
grammars}, for example, simplified C 
with 44 NT symbols and 104 production rules, with
a satisfactory result. The comparison of such grammars takes
about 81ms of CPU-time\footnote{On a 5 years old i5 Linux machine with 4 Gbytes RAM.}.

\section{Conclusion}

In this paper we addressed the problem of context free grammars comparison
from the mathematical analysis point of view. A substitution of terminal letters
by  matrices allows us to reduce the comparison problem to numerical
solution of systems of nonlinear matrix equations.
Besides the elegance of the process,
 this method constitutes a solid base for 
construction of algorithms and tools for comparison of 
context free grammars.
Our experiments with a built prototype show that 
the use of this comparison method with $(2\times 2)$ and 
$(3\times 3)$-matrices in problems appearing in e-learning framework
and even in cases of large grammars is very efficient.

\bibliographystyle{splncs}
\bibliography{paper3}

\end{document}